\documentclass[epj]{svjour} 

\usepackage{graphicx}
\usepackage{subfigure}
\usepackage{epsfig}
\usepackage{amssymb}
\usepackage{calrsfs}
\usepackage{graphics}
\usepackage{txfonts}
\usepackage{pslatex}

\newcommand{\sqrtsnn}{\sqrt{s_{\mbox{\tiny{\it{NN}}}}}}
\newcommand{\sqrts}{\sqrt{s}}
\def\mean#1{\ensuremath{\left<#1\right>}}

\newcommand{\jpsi}{J/\psi}

\newcommand{\ups}{\Upsilon}

\newcommand{\dNdeta}{dN_{\ensuremath{\it ch}}/d\eta|_{\eta=0}}

\newcommand{\gaga}{\gamma\,\gamma}
\newcommand{\gp}{\gamma\,p}
\newcommand{\gA}{\gamma\,A}

\begin{document}
\title{Low-$x$ QCD with CMS at the LHC}
\author{David d'Enterria\inst{1} for the CMS collaboration}
\institute{$^1$CERN, PH-EP, CH-1211 Geneva 23, Switzerland}


%
\date{Received: date / Revised version: date}
%
\abstract{The physics of gluon saturation and non-linear  evolution at small values of  
parton momentum fraction $x$ in the proton and nucleus is discussed in the context of experimental 
results at HERA and RHIC. The rich physics potential of low-$x$ QCD studies at the LHC 
is discussed and some measurements in pp, pA and AA collisions accessible with the Compact Muon
Solenoid (CMS) experiment are presented.
} 

\maketitle

\section{Introduction} 
\label{sec:intro}

\subsection{Parton structure and evolution}

\begin{figure}
\begin{center}
\epsfig{file=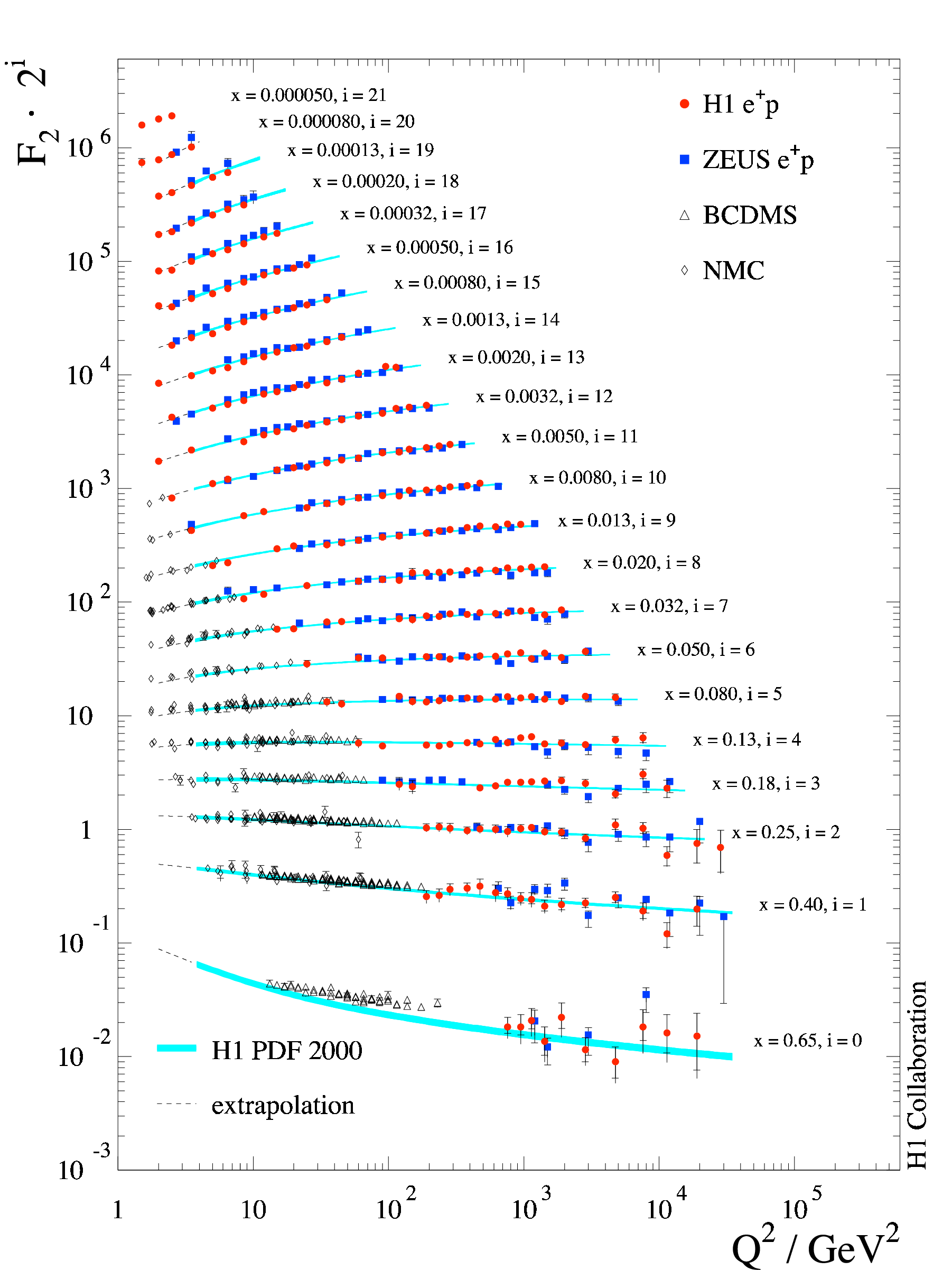,width=7.5cm,height=7.5cm}
\caption{$F_2(x,Q^2)$ measured in DIS at HERA and fixed-target experiments.
\label{fig:HERA_F2}}
\end{center}
\vspace{-.7cm}
\end{figure}
\begin{figure}
\begin{center}
\epsfig{file=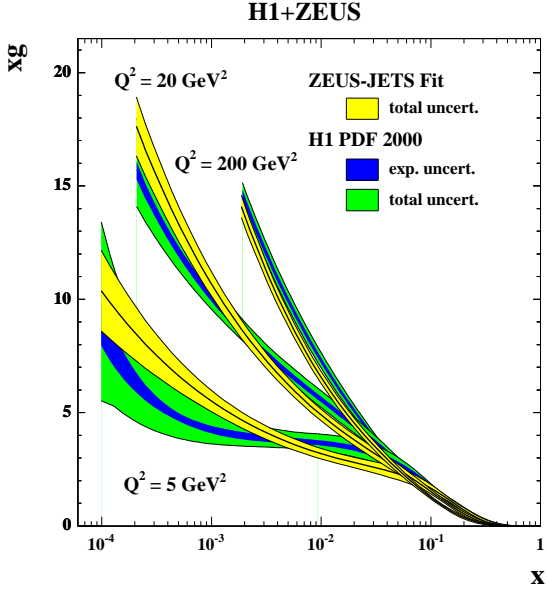, width=0.45\textwidth}
\caption{Gluon distributions extracted at HERA (H1 and ZEUS) as a function 
of $x$ in three bins of $Q^2$~\protect\cite{hera_lhc05}.
\label{fig:HERA_xG}}
\end{center}
\vspace{-.4cm}
\end{figure}

\begin{sloppypar}
Deep inelastic scattering (DIS) electron-proton $ep$ (and electron-nucleus, $eA$) collisions 
provide a precise means to study the
partonic structure of the proton (and nucleus).  
The inclusive DIS hadron cross section, $d^2\sigma/dx\,dQ^2$,  is a function of 
the virtuality $Q^2$ of the exchanged gauge boson  (i.e. its ``resolving power''), and 
the Bjorken-
$x$ fraction of the total nucleon momentum carried by the struck parton. 
The differential cross section 	for the neutral-current ($\gamma,Z$ exchange) process is 
written in terms of the target structure functions as
\begin{equation}
\frac{d^2\sigma}{dx\,dQ^2} = \frac{2\pi\alpha^2}{x\,Q^4}\left[Y_{+}\cdot F_2\mp Y_{\_}\cdot x F_3-\vary^2\cdot F_L\right]\,,
\end{equation}
where $Y_{\pm}=1\pm(1-\vary)^2$ is related to the inelasticity $\vary$ of the collision, and the 
structure functions $F_{2,3,L}(x,Q^2)$ describe the density of quarks and gluons in the hadron\footnote{
$F_2\propto e^2_q\;x\;\Sigma_i(q_i+\bar{q}_i)$, $xF_3\propto x\,\Sigma_i(q_i-\bar{q}_i)$, 
$F_L\propto \alpha_s\,xg$.}. $F_2$ is the dominant contribution to the cross section over most 
of phase space. One of the most significant discoveries at HERA is the strong growth of the 
inclusive DIS cross section 
for decreasing Bjorken-$x$ at fixed $Q^2$ as well as for increasing $Q^2$ at fixed $x$ (Fig.~\ref{fig:HERA_F2}).
The strong scaling violations evident at small $x$ in Fig.~\ref{fig:HERA_F2} are indicative of the 
increasing gluon radiation. At small $x$, $F_2$  is sensitive to the sea quark distribution, driven by
the gluon splitting, and since 
\begin{equation}
\partial F_2(x/2,Q^2)/\partial \ln Q^2 \propto\alpha_s(Q^2)\,xg(x,Q^2)\,,
\label{eq:xG}
\end{equation}
the gluon  density $xg(x,Q^2)$ can be thus determined (Fig.~\ref{fig:HERA_xG}). 
Once measured at an input scale $Q_0^2\gtrsim$ 2 GeV$^2$, 
the parton distribution functions (PDFs) at any other $Q^2$ are given by the 
Dokshitzer-Gribov-Lipatov-Altarelli-Parisi (DGLAP) evolution equations which govern the 
probability of parton branchings (gluon splitting, $q,g-$strahlung) in QCD~\cite{dglap}.
\end{sloppypar}

\begin{sloppypar}
The DGLAP parton evolution, however, only takes into account the $Q^2$-dependence of the PDFs,
resumming over single logarithms in $\alpha_s \ln(Q^2)$, ``leading twist'', but neglecting the $1/x$ terms. 
At large energies (small $x$), the probability of emitting an extra gluon increases as
$\propto\alpha_s\ln(1/x)$. In this regime, the evolution of parton densities proceeds over 
a large rapidity region,  $\Delta y\sim \ln(1/x)$, and the finite transverse momenta 
of the partons become increasingly important. Thus, their appropriate description 
is in terms of {\it $k_T$-unintegrated} PDFs, $xg(x,k_T)$,
described by the Balitski-Fadin-Kuraev-Lipatov (BFKL) equation which governs parton 
evolution in $x$ at fixed $Q^2$~\cite{bfkl}. Hints of extra BFKL radiation have been recently found in 
the enhanced production of forward jets at HERA compared to DGLAP expectations~\cite{hera_forward_jets,marquet05}.
\end{sloppypar}

\subsection{Parton saturation and non-linear evolution at low $x$}

\begin{figure}
\begin{center}
\epsfig{figure=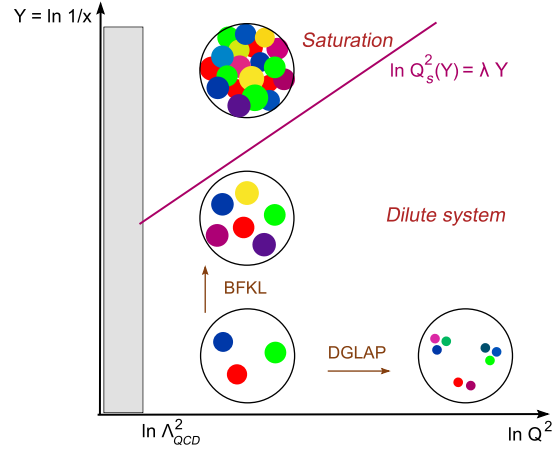, width=0.4\textwidth}
\caption{QCD ``phase diagram'' in the $1/x,Q^2$ plane (each dot represents a parton with 
transverse area $\sim 1/Q^2$ carrying a fraction $x$ of the hadron momentum)~\protect\cite{iancu06}. 
\label{fig:CGC_phase_diag}}
\end{center}
\vspace{-.4cm}
\end{figure}

\begin{sloppypar}
As shown in Figures~\ref{fig:HERA_F2} and~\ref{fig:HERA_xG}, the gluon density rises
very fast for decreasing $x$. For $x<0.01$, the growth in $F_2$ is well described by  
$F_2(x,Q^2)\propto x^{-\lambda(Q^2)}$ with $\lambda \approx$ 0.1 -- 0.3 
logarithmically rising with $Q^2$~\cite{adloff01}. 
Eventually, at some small enough value of $x$ one expects to enter a regime where the gluon 
density becomes so large that non-linear ($gg$ fusion) effects 
become important, taming the growth of the parton densities.
In such a high-gluon density regime three things are expected to occur: 
(i) the standard DGLAP and BFKL {\it linear} equations should no longer be applicable since 
they only account for single parton branchings ($1\rightarrow$2 processes) 
but not for non-linear ($2\rightarrow$1) gluon recombination; 
(ii) pQCD (collinear and $k_T$) factorization should break due to its (now invalid) 
assumption of {\it incoherent} parton scattering; and, as a result, 
(iii) standard pQCD calculations lead to a {\it violation of unitarity} 
even for $Q^2\gg \Lambda^2$. Figure~\ref{fig:CGC_phase_diag}  schematically depicts
the different domains of the parton density as a function of $y=\ln(1/x)$ and $Q^2$. 
The transition to the regime of saturated PDFs is expected for small $x$ values below 
an energy-dependent ``saturation momentum'', $Q_s$, intrinsic to the ({\it size} of the) hadron. 
Since $xg(x,Q^2)$ can be interpreted as the number of gluons with transverse 
area $r^2 \sim 1/Q^2$ in the hadron wavefunction, an increase of $Q^2$ effectively diminishes 
the `size' of each parton, partially compensating for the growth in their number 
(i.e. the higher $Q^2$ is, the smaller the $x$ at which saturation sets in). 
Saturation effects are, thus, expected to occur when the size occupied by the partons 
becomes similar to the size of the hadron, $\pi R^2$. This provides a definition for 
the saturation scale of an arbitrary hadron with $A$ nucleons (i.e. with gluon 
density $xG=A\cdot xg$):
\begin{equation}
Q_s^2(x)\simeq \alpha_s \frac{1}{\pi R^2}\,xG(x,Q^2)\sim A^{1/3}\,x^{-\lambda} \sim A^{1/3}(\sqrts)^{\lambda} \sim A^{1/3}e^{\lambda y},
\label{eq:Qs}
\end{equation}
with $\lambda\approx$ 0.25~\cite{kharzeev_kln}. Equation~(\ref{eq:Qs}) indicates that $Q_s$ 
grows with the number of nucleons, $A$, of the target, and 
the energy of the collision, $\sqrts$, or equivalently, the rapidity of the gluon $y=\ln(1/x)$.
The mass number, $A$, dependence implies that, at equivalent energies, saturation effects will be
enhanced by factors as large as $A^{1/3}\approx$ 6 in heavy nuclear targets ($A$ = 208 for Pb)
compared to protons.
In the last fifteen years, an effective field theory of QCD in the high-energy (high density, small $x$) 
limit has been developed - the Colour Glass Condensate (CGC)~\cite{cgc} - which describes the 
hadrons in terms of classical fields (saturated gluon wavefunctions) below the saturation scale $Q_s$. 
The saturation momentum $Q_s$ introduces a (semi-)hard scale, $Q_s\gg\Lambda$, 
which not only serves as an infrared cut-off to unitarize the cross sections but allows weak-coupling 
perturbative calculations ($\alpha_s(Q_s)\ll$1) in a strong $F_{\mu\nu}$ colour field background.
Hadronic and nuclear collisions are seen as collisions of classical wavefunctions which ``resum'' all gluon 
recombinations and multiple scatterings.
The quantum evolution in the CGC approach is given by the JIMWLK~\cite{jimwlk} non-linear equations
(or by their mean-field limit for $N_c\rightarrow\infty$, the Balitsky-Kovchegov equation~\cite{bk}) 
which reduce to the standard BFKL kernel at higher $x$ values.
\end{sloppypar}


\section{Parton saturation: Experimental studies}
\label{sec:signatures}

\begin{sloppypar}
The main  source of information on the {\it quark} densities is obtained from measurements of 
(i) the structure functions $F_{2,3}$ in lepton-hadron scattering, and 
(ii) lepton pair (Drell-Yan) production in hadron-hadron collisions. 
The {\it gluon} densities, $xG$ enter at LO directly  in hadron-hadron scattering processes with 
(i) prompt photons and (ii) jets in the final state, as well as in the (difficult) measurement of 
(iii) the longitudinal DIS structure function $F_L$ (and also indirectly in $F_2$ through the 
derivative in Eq.~(\ref{eq:xG})). In addition, (iv) heavy vector mesons ($\jpsi,\ups$) from
diffractive photoproduction processes\footnote{Diffractive $\gp$ ($\gA$) processes are characterized 
by a quasi-elastic interaction - mediated by a Pomeron or two gluons in a colour singlet state -
in which the p (A) remains intact (or in a low excited state) and separated by a rapidity 
gap from the rest of final-state particles.} 
are a valuable probe of the gluon density since their cross sections are proportional to 
the {\it square} of $xG$~\cite{ryskin95,teubner05}:
\begin{eqnarray}
\left .\frac{d\sigma_{\gp,A\rightarrow V\,p,A}}{dt}\right|_{t=0} =
\frac{\alpha_s^2\Gamma_{ee}}{3\alpha M_V^5}16\pi^3\left[xG(x,Q^2)\right]^2\,, \label{eq:diffract_qqbar_sigma}\\
\;\;\mbox{with }\;Q^2=M_V^2/4\;\;\mbox{ and }\;x=M_V^2/W_{\gp,A}^2.
\label{eq:diffract_qqbar_x}
\end{eqnarray}

In hadronic collisions, one commonly measures (real and virtual) photons and jets at 
central rapidities ($y=0$) where $x=x_T=Q/\sqrts$, with $Q\sim p_T,M$  the characteristic scale 
of the hard scattering. However, one can probe smaller $x_2$ values in the target by measuring 
the corresponding cross sections in the {\it forward} direction. Indeed, for a $2\rightarrow 2$ 
parton scattering the {\it minimum} momentum fraction probed in a process with a particle 
of momentum $p_T$ produced at pseudo-rapidity $\eta$ is~\cite{vogels_dAu}
\begin{equation}
x_{2}^{min} = \frac{x_T\,e^{-\eta}}{2-x_T\,e^{\eta}}\;\; \mbox{ where } \;\; x_T=2p_T/\sqrt{s}\,,
\label{eq:x2_min}
\end{equation}
i.e. $x_2^{min}$ decreases by a factor of $\sim$10 every 2 units of rapidity. Though Eq.~(\ref{eq:x2_min}) 
is a lower limit at the end of phase-space (in practice the $\mean{x_2}$ values in parton-parton
scatterings are at least 10 larger than $x_2^{min}$~\cite{vogels_dAu}), it provides the right 
estimate of the typical $x_2=(p_T/\sqrts)\,e^{\eta}$ values reached in non-linear $2\rightarrow 1$ 
processes (in which the momentum is balanced by the gluon ``medium'') as described in parton saturation 
models~\cite{accardi04}.
\end{sloppypar}

\begin{figure}
\begin{center}
\epsfig{file=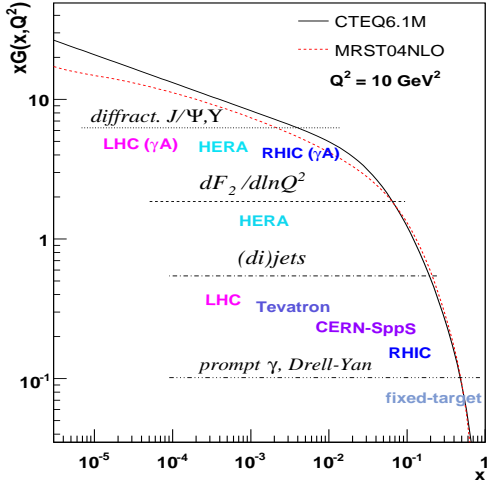,width=6.8cm,height=6.8cm}
\end{center}
\vspace{-0.4cm}
\caption{Experimental measurements at various facilities providing information 
on the gluon PDF in different ranges of Bjorken-$x$.}
\label{fig:xG_processes}
\end{figure}

\begin{figure}[htb]
\begin{center}
\epsfig{file=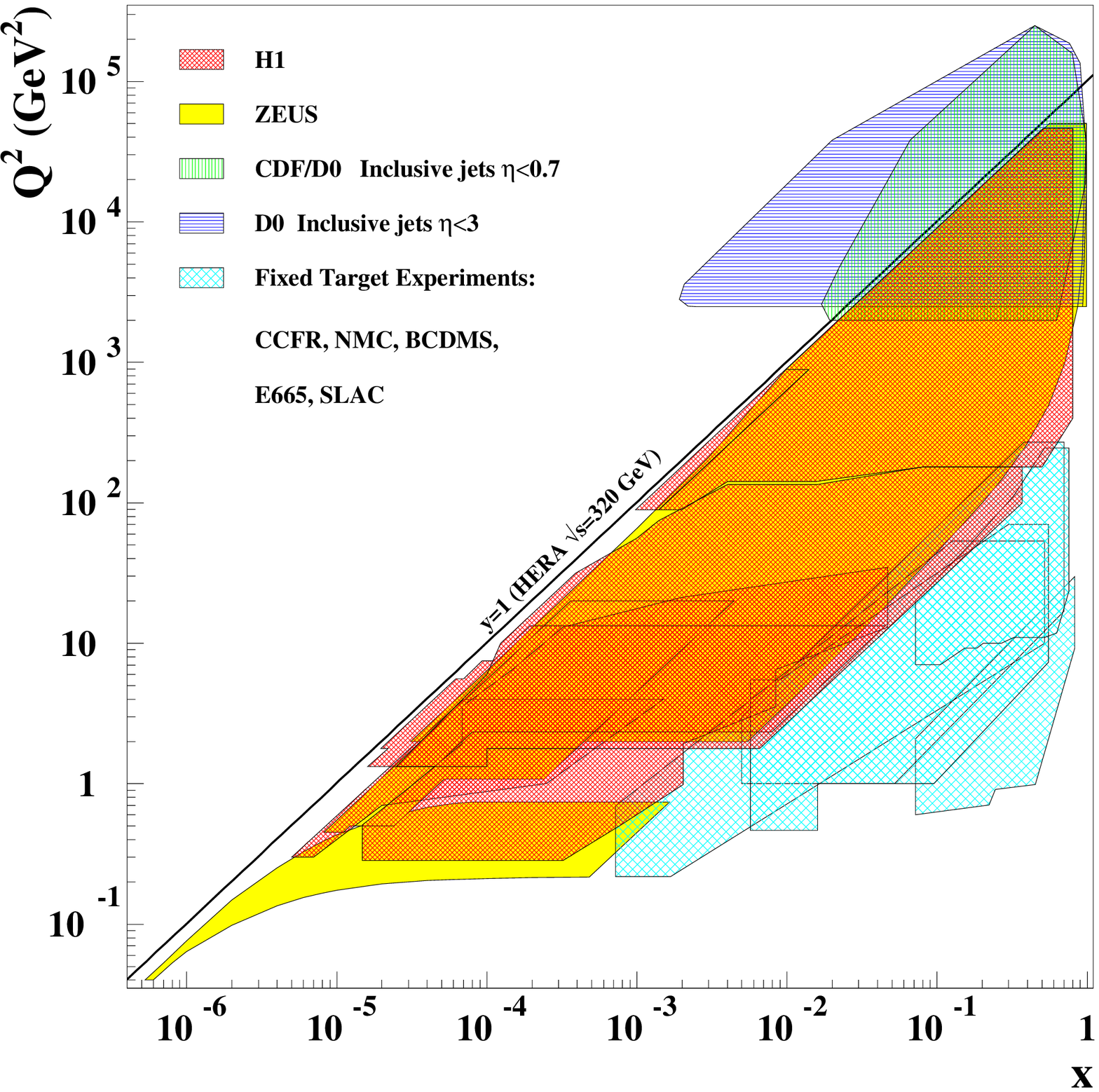,width=7.cm,height=5.cm}
\hspace*{-.3cm}
\epsfig{file=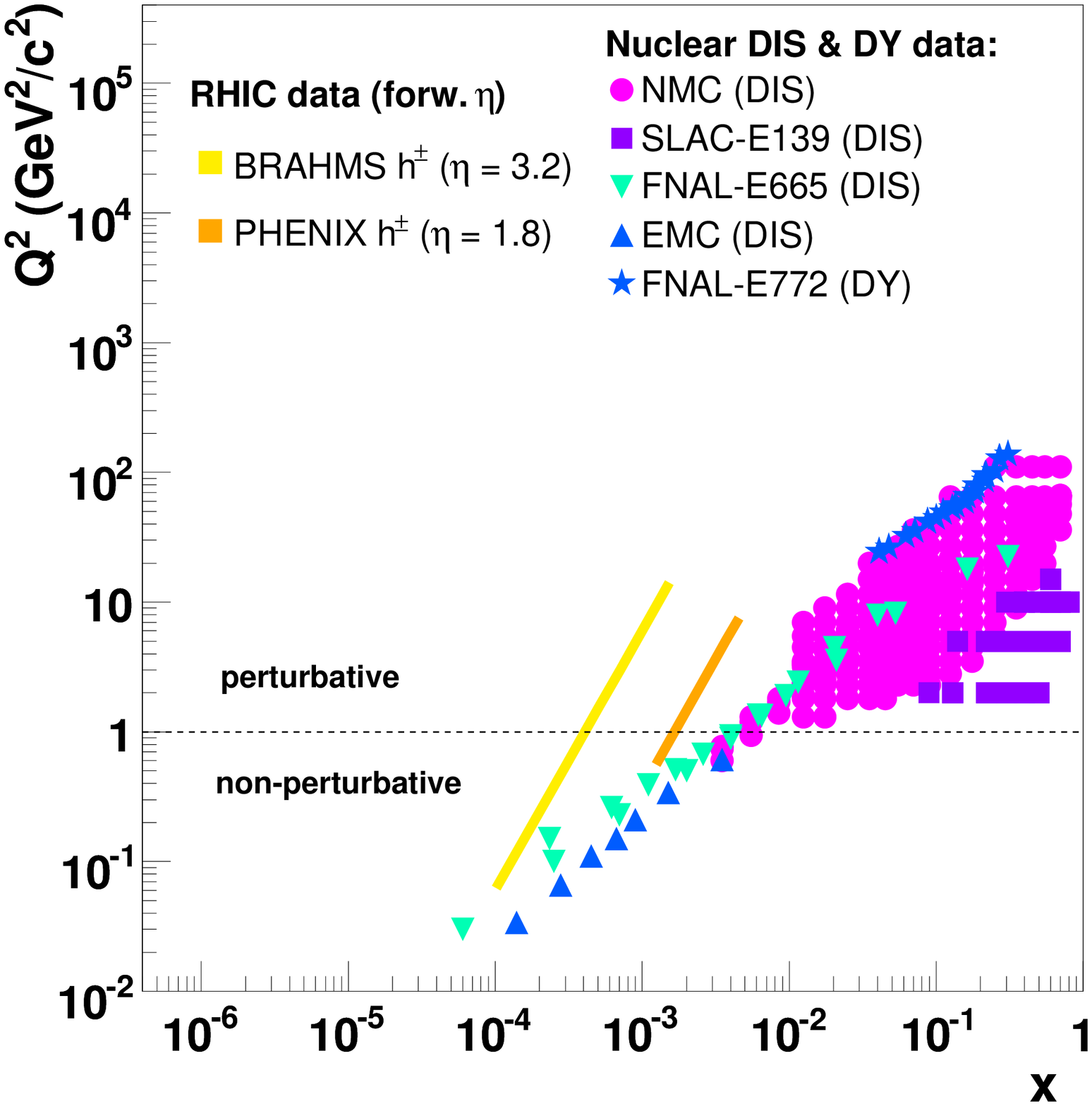,width=7.1cm,height=5.2cm}
\caption{Available measurements in the  $(x,Q^2)$ plane  used for the 
determination of the proton~\protect\cite{newman03} (top) and nuclear~\protect\cite{dde_qm04} (bottom) PDFs.}
\label{fig:x_Q2_map}
\end{center}
\vspace{-0.4cm}
\end{figure}

\begin{figure}[htb]
\begin{center}
\epsfig{file=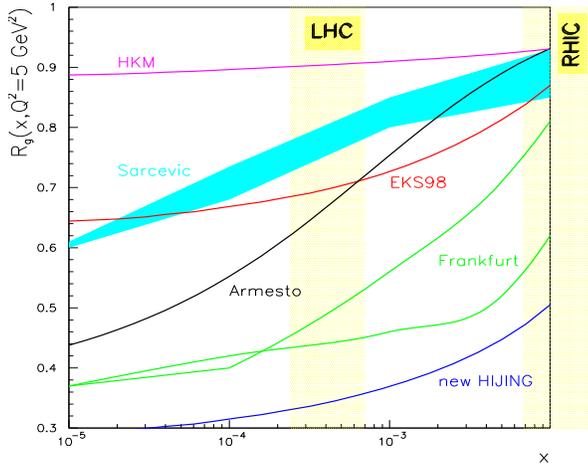,width=8.cm,height=7.cm}
\caption{Ratios of the Pb over proton gluon PDFs versus $x$ from different models
at $Q^2$ = 5 GeV$^2$. 
Figure taken from~\protect\cite{yellowrep_pdf}.}
\label{fig:xG_Pb}
\end{center}
\vspace*{-.5cm}
\end{figure}

\begin{sloppypar}
Figure~\ref{fig:xG_processes} summarizes the range of experimental processes sensitive
to the gluon density and their approximate $x$ coverage. Figure~\ref{fig:x_Q2_map} 
shows the kinematical map in $(x,Q^2)$ of the DIS, DY, direct $\gamma$ and jet data used 
in the PDF fits. Results from HERA and the Tevatron cover a substantial range of the proton 
structure ($10^{-4}\lesssim x\lesssim0.8$, 1 $\lesssim Q^2\lesssim$ 10$^{5}$ GeV$^2$) 
but the available measurements are much rarer  in the case of nuclear targets (basically 
limited to fixed-target studies, $10^{-2}\lesssim x\lesssim 0.8$ and
1 $\lesssim Q^2\lesssim$ 10$^{2}$ GeV$^2$). As a matter of fact, the nuclear parton distributions 
are basically unknown at low $x$ ($x<0.01$) where the only available measurements are
fixed-target data in the {\it non-perturbative} range ($Q^2<1$ GeV$^2$) dominated by
Regge dynamics rather than quark/gluon degrees of freedom. An example of the current
lack of knowledge of the nuclear densities at low $x$ is presented in Fig.~\ref{fig:xG_Pb}
where different available parametrizations of the ratio of Pb to proton gluon distributions,
consistent with the available nDIS data at higher $x$, show differences as large as a factor 
of three~\cite{armesto_shadow}.
\end{sloppypar}

\subsection{HERA results}
\label{sec:hera}

\begin{figure}
\begin{center}
\hspace*{.5cm}
\centerline{\epsfig{file=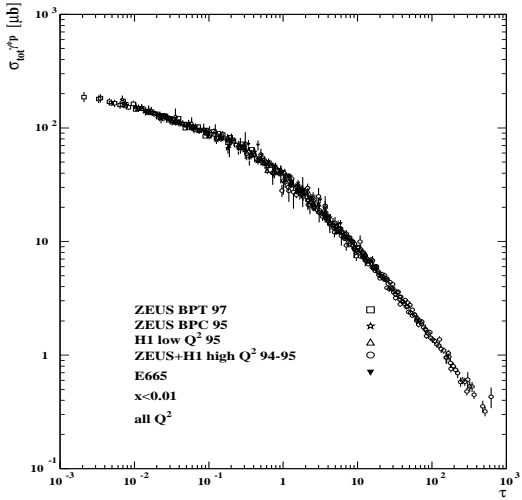,width=8.7cm,height=7.cm}}
\caption{Geometric scaling in the DIS $\gamma^\star p$ cross sections plotted 
versus $\tau=Q^2/Q_s^2$ in the range $x < 0.01$, 
0.045 $< Q^2 <$ 450 GeV$^2$~\protect\cite{golec_biernat_wusthoff}.}
\label{fig:geom_scaling}
\end{center}
\vspace*{-.4cm}
\end{figure}

\begin{sloppypar}
Though the large majority of $ep$ DIS data collected during the HERA-I phase are consistent
with standard DGLAP predictions, 
more detailed and advanced experimental and theoretical results in the recent years have pointed 
to interesting hints of non-linear QCD effects in the data. Arguably, the strongest manifestation of 
such effects is given by the so-called ``geometric scaling'' property observed in 
inclusive $\sigma_{DIS}$ for $x<0.01$~\cite{golec_biernat_wusthoff}
as well as in various diffractive cross sections~\cite{forshaw,marquet06}. For inclusive 
DIS events, this feature manifests itself in a total cross section at small $x$ ($x<0.01$) which 
is only a function of  $\tau\!=\!Q^2/Q_s^2(x)$, instead of being a function 
of $x$ and $Q^2/Q_s^2$ separately (Fig.~\ref{fig:geom_scaling}). The saturation momentum
follows $Q_s(x)=Q_0(x/x_0)^{\lambda}$ with parameters $\lambda\sim$ 0.3, $Q_0$ = 1 GeV, 
and $x_0\sim$ 3$\cdot$10$^{-4}$. 
Interestingly, the scaling is valid up to very large values of $\tau$, well above the saturation scale, 
in an ``extended scaling'' region with $Q_s^2<Q^2<Q_s^4/\Lambda_{\ensuremath{\it QCD}}^2$~\cite{ext_scal,iancu06}. 
The saturation formulation is suitable to describe not only inclusive DIS, but also inclusive diffraction 
$\gamma^\star p \rightarrow X\,p$. The very similar energy dependence of the inclusive diffractive 
and the total cross section in $\gamma^\star p$ collisions at a given $Q^2$ is easily explained 
in the Golec-Biernat W\"usthoff model~\cite{golec_biernat_wusthoff} but not in standard
collinear factorization. Furthermore, geometric scaling has been also found in different diffractive
DIS cross sections (inclusive, vector mesons, deeply-virtual Compton scattering DVCS)~\cite{forshaw,marquet06}. 
All these results suggest that the observed scalings are indeed manifestations of the saturation
regime of QCD. Unfortunately, the value of $Q_s\sim$ 1 GeV at HERA lies in the transition 
region between the soft and hard sectors and, therefore, non-perturbative effects obscure the 
obtention of clearcut experimental signatures.
\end{sloppypar}

\subsection{RHIC results}
\label{sec:rhic}

\begin{sloppypar}
The expectation, based on Eq.~(\ref{eq:Qs}), of enhanced parton saturation effects in the nuclear wave
functions accelerated at ultra-relativistic energies has been one of the primary physics motivations
for the heavy-ion programme at RHIC\footnote{The saturation scale at $y$ =0 in Au at RHIC is $Q_s^2\sim$ 2 GeV, 
much larger than that of protons probed at HERA, $Q_s^2\sim$ 0.5 GeV.}~\cite{cgc}.
Further, the properties of the high-density matter produced in the final-state of AA interactions cannot be properly 
interpreted without having determined the influence of {\it initial state} modifications of the nuclear PDFs.
In this context, after five years of operation, two main experimental observations at RHIC have been found 
consistent with CGC predictions: (i) the modest hadron multiplicities measured in AuAu reactions, and 
(ii) the suppressed hadron yield at forward rapidities in dAu collisions.

\begin{figure}[htb]
\begin{center}
\vspace{-0.6cm}
\epsfig{file=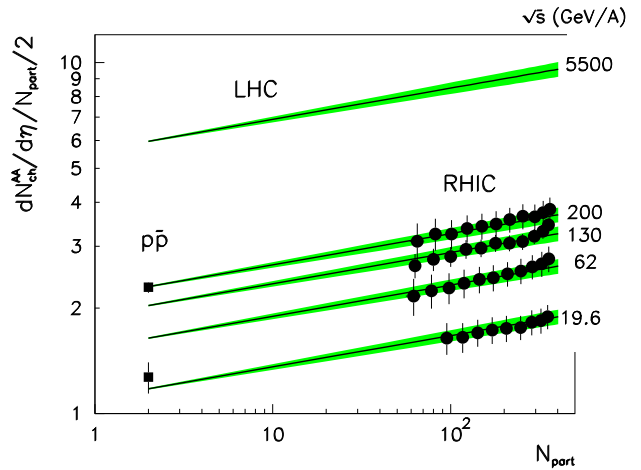,width=8.4cm}
\caption{Dependences on c.m. energy and centrality (given in terms of the number of nucleons participating in the collision, 
$N_{\rm part}$) of  $\dNdeta$  (normalized by $N_{\rm part}$):
PHOBOS AuAu data~\protect\cite{phobos_wp} vs the predictions of the saturation approach~\protect\cite{armesto04}. }
\label{fig:dNdeta}
\end{center}
\vspace{-.4cm}
\end{figure}

The bulk, $\dNdeta\approx$ 700, multiplicities measured at mid-rapidity in central AuAu at 
$\sqrtsnn$ = 200 GeV are comparatively lower than the $\dNdeta\approx$ 1000 predictions~\cite{eskola_qm01} 
of ``minijet'' scenarios, soft Regge models, 
or extrapolations from an incoherent sum of proton-proton collisions, but they can be reproduced 
by approaches based on gluon saturation~\cite{kharzeev_kln,armesto04} which take into account 
a reduced parton flux in the nuclear targets, i.e. $f_{a/A}(x,Q^2)<A\cdot f_{a/N}(x,Q^2)$. 
In the CGC calculations, the final hadron multiplicities are assumed to be simply related to 
the initial number of released partons (local parton-hadron duality) which are depleted in the 
initial state compared to pp collisions due to non-linear gluon-gluon recombinations~\cite{kharzeev_kln}. 
Simple assumptions related to the dependence of the saturation scale on energy and overlapping 
area of the colliding nuclei, describe the centrality and center-of-mass (c.m.) energy dependences of the 
bulk AA hadron production (Fig.~\ref{fig:dNdeta}).
\end{sloppypar}

\begin{figure}[htb]
\begin{center}
\includegraphics[width=7.8cm,height=6.cm]{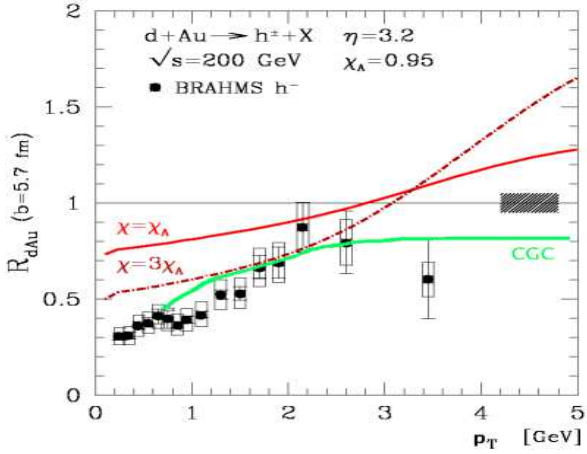}
\caption{Nuclear modification factor $R_{dAu}(p_T)$ for charged hadrons produced in dAu  
at $\sqrtsnn$ = 200 GeV: BRAHMS data~\protect\cite{brahms_wp} versus DGLAP 
shadowing~\protect\cite{accardi04} and CGC~\protect\cite{tuchin04} predictions. 
Fig. adapted from~\protect\cite{accardi04}.}
\label{fig:RdA}
\end{center}
\vspace{-.4cm}
\end{figure} 

\begin{sloppypar}
The second manifestation of saturation-like effects in the RHIC data is the BRAHMS observation~\cite{brahms_wp} 
of suppressed yields of moderately high-$p_T$ hadrons ($p_T\approx 2 - 4 $ GeV/$c$) in dAu relative
to pp collisions at $\eta\approx$ 3.2. Hadron production at such small angles is 
sensitive to partons in the Au nucleus with $x\approx$ $\mathcal{O}$(10$^{-3}$).
The observed nuclear modification factor, $R_{dAu}\approx$ 0.8, cannot be reproduced by pQCD
calculations that include standard {\it leading-twist} shadowing of the nuclear PDFs~\cite{accardi04,vogels_dAu}
but can be described by CGC approaches~\cite{tuchin04} that parametrize the Au nucleus as a saturated 
gluon wavefunction. In addition, a recent analysis of the nuclear DIS $F_2$ data also confirms the existence 
of ``geometrical scaling'' for $x<$0.017~\cite{armesto04}.
\end{sloppypar}



\section{Low-$x$ QCD at the LHC}
\label{sec:lhc}

\begin{figure}[htb]
\begin{center}
\centerline{\epsfig{file=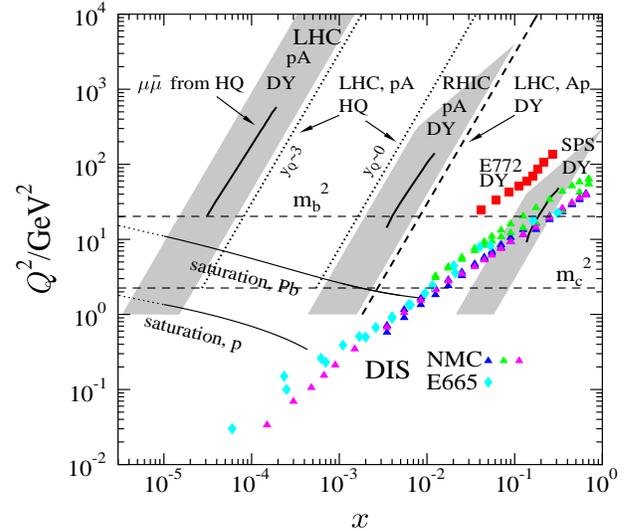,width=8.cm,height=7.cm}}
\caption{Kinematical $(x,Q^2)$ range probed 
at various rapidities $y$ and c.m. energies in $\sqrtsnn$ = 8.8 TeV pA collisions at the LHC~\protect\cite{yellowrep_pdf}.}
\label{fig:xQ2_map_pA_LHC}
\end{center}
\vspace*{-.5cm}
\end{figure}

\begin{sloppypar}
The Large Hadron Collider (LHC) at CERN will provide pp, pA and AA collisions 
at $\sqrtsnn$ = 14, 8.8 and 5.5 TeV respectively with 
luminosities $\mathcal{L}\sim$  10$^{34}$, 10$^{29}$ and 5$\cdot$10 $^{26}$ cm$^{-2}$ s$^{-1}$.
Such large c.m. energies and luminosities will allow detailed QCD studies at unprecedented 
low $x$ values thanks to the copious production of hard probes (jets, quarkonia, 
prompt $\gamma$, Drell-Yan pairs, etc.). 
The expected advance in the study of low-$x$ QCD phenomena will be specially substantial 
for nuclear systems since the saturation momentum, Eq.~(\ref{eq:Qs}),  
$Q_s^2 \approx$ 5 -- 10 GeV$^2$, will be in the perturbative range~\cite{kharzeev_kln}, and
the relevant $x$ values, Eq.~(\ref{eq:x2_min}), will be 30--70 times lower than 
AA and pA reactions at RHIC: $x \approx 10^{-3}(10^{-5})$ at central (forward) 
rapidities for processes with $Q^2\sim$10 GeV$^2$ (Fig.~\ref{fig:xQ2_map_pA_LHC}).
\end{sloppypar}

\subsection{The CMS experiment}
\label{sec:cms}

\begin{sloppypar}
The CMS experiment is one of the two large general-purpose detectors being installed at the LHC. 
Its experimental capabilities are extremely well adapted for the study of low-$x$ phenomena 
with proton and ion beams featuring:
\begin{description}
\item (i) Very large acceptance at midrapidity ($|\eta|<2.5$, full $\phi$) for 
charged and neutral hadrons as well as $\mu^\pm$, $e^\pm$, and $\gamma$ over a wide range of $p_T$
(the 4 T magnetic field results in the best track momentum resolution at LHC).
\item (ii) Excellent muon reconstruction leading to the best mass resolution for 
$J/\psi$, and $\Upsilon$ measurements at the LHC. 
\item (iii) Complete electromagnetic (EM) and hadronic (HAD) calorimetry for full jet 
reconstruction over $|\eta|<3$ and $\Delta\phi$ = 2$\pi$ with a large statistical significance for single
jet and jet+$X$ ($X$ = jet, $\gamma$, $Z$) channels.
\item (iv) Unparalleled forward physics capabilities thanks to the forward hadronic calorimeter 
(HF, 3$<|\eta|<$5), TOTEM T1 (3.1 $<|\eta|<$ 4.7) and T2 (5.5$<|\eta|<$6.6)
trackers, and CASTOR (5.3$<|\eta|<$6.7) and zero-degree (ZDC, $|\eta|>$8.1 for neutrals) calorimeters.
\end{description}

\begin{figure}[htb]
\begin{center}
\hspace*{.3cm}
\centerline{\epsfig{file=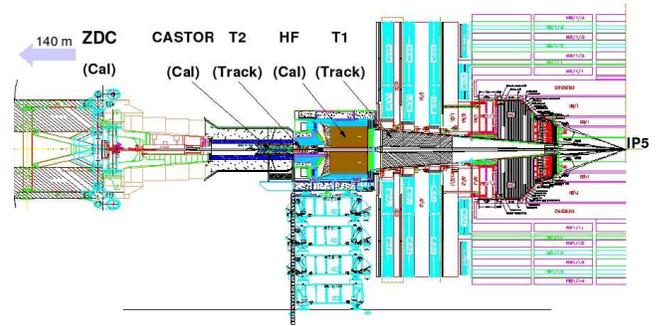,width=0.47\textwidth}}
\caption{Layout of the detectors in the CMS forward region.}
\label{fig:forw_CMS}
\end{center}
\vspace*{-.8cm}
\end{figure}
The combination of HF, TOTEM, CASTOR and ZDC (Fig.~\ref{fig:forw_CMS}) makes of CMS 
the largest acceptance detector ever built at a hadron collider.
The HF~\cite{hf}, located 11.2 m away on both sides of the interaction point (IP), is a steel plus quartz-fiber 
\v{C}erenkov calorimeter with 1200 channels ($\Delta\eta\times\Delta\phi\sim$0.18$\times$0.18, 1.65 m
absorber corresponding to 10.3$\lambda_I$) sensitive to the deposited EM and HAD energy, 
allowing jet reconstruction at very forward rapidities. The T1 and T2 telescopes are part of the the 
TOTEM experiment~\cite{totem} which shares the same IP as CMS and are mainly designed to measure charged 
tracks from diffractive dissociation processes.
CASTOR~\cite{castor} is an azimuthally symmetric electromagnetic/hadronic calorimeter situated at 14.37 m from 
the interaction point covering the same acceptance as T2. The calorimeter is a \v{C}erenkov-light device, 
consisting of successive layers of tungsten absorber and fused silica (quartz) plates as active medium
arranged in 
2 EM (10 HAD) sections of about 22$X_0$ (10.3$\lambda_I$) radiation (interaction) lengths.
The ZDC~\cite{zdc} is also a tungsten+quartz sampling \v{C}erenkov calorimeter with 5 EM 
(19$X_0$, divided in $x$) and 4 HAD (5.6$\lambda_I$, divided $z$) sections. It is located at 140 m 
from the CMS vertex at the end of the straight sections of the two LHC pipes containing the 
countercirculating beams. The purpose of the ZDC is to measure very forward going neutrons and 
photons with $\sim$10\% (2 mm) energy (position) resolution.
\end{sloppypar}

\subsection{Low-$x$ QCD measurements in CMS}
\label{sec:cms_lowx}
The following three measurements in pp, pA and AA collisions are being considered in CMS 
to look for signatures of high gluon density effects at low $x$.

\subsubsection{Forward jets (pp, pA, AA)}
\begin{figure}[htb]
\begin{center}
\centerline{\epsfig{file=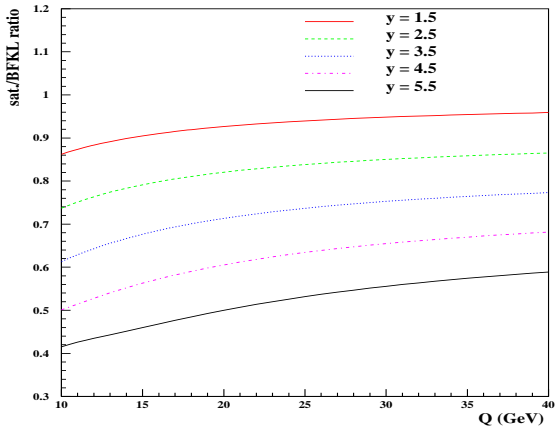,width=7.5cm,height=5.8cm}}
\caption{Ratio of the saturation over BFKL predictions for the M\"uller-Navelet forward
jet cross sections in pp collisions at $\sqrts$ = 14 TeV as a function of $Q\equiv Q_1 = Q_2$ 
for different values of $y\equiv y_1 = -y_2$~\protect\cite{marquet05}.}
\label{fig:muller_navelet_jets}
\end{center}
\vspace*{-.8cm}
\end{figure}

\begin{sloppypar}
The cross section for dijet production in the forward direction, ``M\"uller-Navelet'' jets~\cite{mueller_navelet}, 
is a particularly sensitive measure of the small $x$ parton dynamics in hadronic collisions~\cite{marquet05}.
The two HF calorimeters (3$<|\eta|<$5), specifically designed to measure energetic forward jets\footnote{The 
HF plays a prominent role in forward jet tagging for the vector-boson-fusion 
($qq\rightarrow qqH$) Higgs production channel.}, have an energy resolution of  $\sim$20\%
for typical jets with $E_T\sim$ 40 GeV (i.e. $E = E_T\,\cosh\eta\approx$ 1 TeV at $\eta$= 4).
In the presence of low-$x$ saturation effects, the forward-backward dijet production cross section
(separated by $\Delta\eta\sim$ 9 and, thus, measurable in each of the HFs) is expected to be suppressed by a
factor of $\sim$2 in pp at 14 TeV (Fig.~\ref{fig:muller_navelet_jets}). A study is underway to 
determine the feasibility of such measurements in CMS.
\end{sloppypar}

\subsubsection{$Q\bar{Q}$ photoproduction (electromagnetic AA collisions)}
\label{sec:upc_cms}
\begin{sloppypar}
High-energy diffractive production of heavy vector mesons ($\jpsi,\ups$) proceeds through colourless 
two-gluon exchange (which couples to $\gamma\rightarrow Q\bar{Q}$) and is thus a sensitive 
probe of the low-$x$ gluon densities, see Eq.~(\ref{eq:diffract_qqbar_sigma}).
Ultra-peripheral interactions (UPCs) of high-energy heavy ions generate strong electromagnetic fields 
which help constrain the low-$x$ behaviour of $xG$ via quarkonia produced in $\gamma$-nucleus 
collisions~\cite{dde_qm05}. Lead beams at 2.75 TeV have Lorentz factors $\gamma$ = 2930 
leading to maximum (equivalent) photon energies $\omega_{\ensuremath{\it max}}\approx \gamma/R\sim$ 100 GeV,
and c.m. energies 
$W_{\gaga}^{\ensuremath{\it max}}\approx$ 160 GeV and $W^{\ensuremath{\it max}}_{\gA}\approx$ 1 TeV. 
From Eq.~(\ref{eq:diffract_qqbar_x}), the $x$ values probed in $\gA\rightarrow\jpsi \;A$ processes 
at $y$ = 2 can be as low as $x\sim 10^{-5}$. 
The CMS experiment can measure $\ups\rightarrow \mu^+\mu^-$ produced in
electromagnetic PbPb collisions tagged with neutrons detected in the ZDCs
(as done at RHIC~\cite{dde_qm05}). Figure 13 shows the 
expected dimuon invariant mass distributions predicted by {\sc starlight}~\cite{starlight} 
within the CMS acceptance for an integrated PbPb luminosity of 0.5 nb$^{-1}$. 
An $\ups$ peak with $\sim$1200 counts is clearly seen on top of the $\mu^+\mu^-$ 
continuum.
\end{sloppypar}

\begin{figure}[htb]
\begin{center}
\epsfig{file=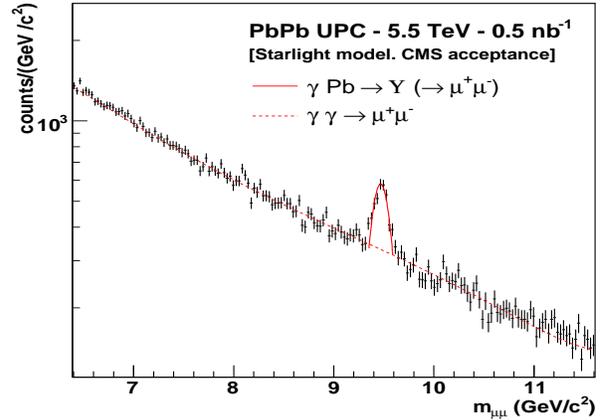,width=8.cm,height=6.cm}
\caption{Expected $\mu^+\mu^-$ invariant mass from $\gamma\,Pb\rightarrow \Upsilon\,Pb^\star\rightarrow \mu^{+}\mu^{-}\,Pb^\star$ 
and $\gaga\rightarrow \mu^+\mu^-$ as given by {\it Starlight}~\protect\cite{starlight} 
for UPC PbPb collisions at $\sqrtsnn$ = 5.5 TeV in the CMS acceptance.}
\end{center}
\label{fig:upc_lhc_starlight}
\vspace*{-.8cm}
\end{figure}
\subsubsection{Forward Drell-Yan pairs (pp, pA, AA)}
\begin{sloppypar}
High-mass Drell-Yan pair production at the very forward rapidities covered by CASTOR and T2  ($|\eta|\sim 5 - 6$)
can probe the parton densities down to $x\sim 10^{-6}$. A study is underway in CMS 
to combine the CASTOR electromagnetic energy measurement together with the good position resolution of T2
for charged tracks, to trigger on and reconstruct the $e^+e^-$ invariant mass in pp collisions at 14 TeV,
and perform a two-dimensional study of $xg$ in the $M^2$ and $x$ plane.
\end{sloppypar}

\section{Conclusion}
\label{sec:conclusion}

\begin{sloppypar}
We have reviewed the physics of non-linear QCD and high gluon densities at small fractional momenta 
$x$  with emphasis on the existing data at HERA (proton) and RHIC (nucleus) which support the 
existence of a parton saturation regime (also known as colour-glass-condensate). The future perspectives 
at the LHC have been presented, including the promising capabilities of the forward CMS detectors to study 
the parton densities down to $x\sim 10^{-6}$ with various hard probes (jets, quarkonia, Drell-Yan). 
The programme of investigating the dynamics of low-$x$ QCD is not only appealing in its own right but 
it is an essential prerequisite for predicting a large variety of hadron-, photon- and neutrino- scattering 
cross sections at very high energies.
\end{sloppypar}

\section*{Acknowledgments}
The author acknowledges valuable comments from R.~Vogt and J.~Jalilian-Marian, and 
thanks J.~Nystrand for providing the {\sc starlight} photoproduction cross sections in 
PbPb at LHC.
This work is supported by the 6th EU Framework Programme contract MEIF-CT-2005-025073.



\begin{thebibliography}{}

\def\IJMPA{{Int. J. Mod. Phys.}~{\bf A}}
\def\EPJ{{Eur. Phys. J.}~{\bf C}}
\def\JPG{{J. Phys.}~{\bf G}}
\def\JHEP{{J. High Energy Phys.}~}
\def\NCA{Nuovo Cimento~}
\def\NIM{Nucl. Instrum. Methods~}
\def\NIMA{{Nucl. Instrum. Methods}~{\bf A}}
\def\NPA{{Nucl. Phys.}~{\bf A}}
\def\NPB{{Nucl. Phys.}~{\bf B}}
\def\PLB{{Phys. Lett.}~{\bf B}}
\def\PLC{Phys. Repts.\ }
\def\PRL{Phys. Rev. Lett.\ }
\def\PRD{{Phys. Rev.}~{\bf D}}
\def\PRC{{Phys. Rev.}~{\bf C}}
\def\ZPC{{Z. Phys.}~{\bf C}}

\bibitem{hera_lhc05} M.~Dittmar {\it et al.}, in {\it Proceeds. HERA and the LHC}, hep-ph/0511119. 
\bibitem{dglap}V.N.~Gribov and L.N.~Lipatov, Sov.\ Journ.\ Nucl.\ Phys.\ {\bf 15} (1972) 438; 
G. Altarelli and G. Parisi, Nucl.\ Phys.\,{\bf B126} (1977) 298; 
Yu. L.~Dokshitzer, Sov.\ Phys.\ JETP {\bf 46} (1977) 641.
\bibitem{bfkl}L.N.~Lipatov, Sov.\ J.\ Nucl.\ Phys.\,{\bf 23} (1976) 338; 
E.A.~Kuraev,  L.N.~Lipatov and V.S.~Fadin, Zh. Eksp. Teor. Fiz {\bf 72},  (1977) 3;  
Ya.Ya.~Balitsky, L.N.~Lipatov, Sov.\ J.\ Nucl.\ Phys. {\bf 28} (1978) 822.
\bibitem{hera_forward_jets}S.~Chekanov {\it et al.}  [ZEUS Collab.],  Phys.\ Lett.\ B {\bf 632} (2006) 13;  
A.~Aktas {\it et al.}  [H1 Collab.], Eur.\ Phys.\ J.\ C {\bf 46} (2006) 27.  
\bibitem{marquet05}C.~Marquet and C.~Royon,  Nucl.\ Phys.\ B {\bf 739} (2006) 131. 
\bibitem{adloff01}C.~Adloff {\it et al.}  [H1 Collab.],  Phys.\ Lett.\ B {\bf 520} (2001) 183.  
\bibitem{iancu06}E.~Iancu, hep-ph/0608086.  
\bibitem{kharzeev_kln}D.~Kharzeev and M.~Nardi, Phys.\ Lett.\ B {\bf 507} (2001) 121; 
D.~Kharzeev, E.~Levin and M.~Nardi, Nucl.\ Phys.\ A {\bf 747} (2005) 609. 
\bibitem{cgc}See e.g. E.~Iancu and R.~Venugopalan, in {\it QGP. Vol 3} Eds: R.C. Hwa and X.N. Wang, World Scientific, Singapore,  
hep-ph/0303204;  
J.~Jalilian-Marian and Y.~V.~Kovchegov,  Prog.\ Part.\ Nucl.\ Phys.\  {\bf 56} (2006) 104; and refs. therein. 
\bibitem{jimwlk}J.~Jalilian-Marian, A.~Kovner, A.~Leonidov and H.~Weigert, \NPB {\bf 504} (1997) 415; 
Phys.\ Rev.\ {\bf D59} (1999) 014014;
E.~Iancu, A.~Leonidov and L.~McLerran, Nucl. Phys.\ {\bf A692} (2001) 583;
\bibitem{bk}I. Balitsky, Nucl. Phys. {\bf B} 463, (1996) 99; Yu.V. Kovchegov, Phys. Rev. {\bf D 61}, (2000) 074018.
\bibitem{ryskin95}M.~G.~Ryskin {\em et al.}, 
\ZPC 76, (1997) 231.
\bibitem{teubner05}T.~Teubner, Proceeds. DIS'05, AIP Conf.\ Proc. {\bf 792} (2006) 416.
\bibitem{vogels_dAu}V.Guzey, M.Strikman, W.Vogelsang, Phys. Lett. B{\bf 603} (2004) 173
\bibitem{accardi04}A.~Accardi, Acta Phys.Hung. A22 (2005) 289. 
\bibitem{newman03}P.~Newman,  Int.\ J.\ Mod.\ Phys.\ A {\bf 19} (2004) 1061.  
\bibitem{dde_qm04}D.~d'Enterria, J.\ Phys.\ G {\bf 30} (2004) S767.
\bibitem{yellowrep_pdf}A.~Accardi {\em et al.}, in CERN Yellow report on {\it Hard probes in Heavy Ion collisions at the LHC}, hep-ph/0308248.
\bibitem{armesto_shadow}N.~Armesto, \JPG \ to appear, hep-ph/0604108. 
\bibitem{golec_biernat_wusthoff}K.~Golec-Biernat and M.~W\"usthoff, Phys.\ Rev. {\bf D59}, (1999) 014017; 
Phys.\ Rev. {\bf D60}, (1999) 114023.
\bibitem{forshaw}J.~R.~Forshaw and G.~Shaw, JHEP {\bf 0412} (2004) 052. 
\bibitem{marquet06}C.~Marquet, L.~Schoeffel, hep-ph/0606079.
\bibitem{ext_scal}E. Iancu, K. Itakura, L. McLerran, \NPA {\bf 708} (2002) 327.
\bibitem{eskola_qm01} K.~J.~Eskola,  Nucl.\ Phys.\ A {\bf 698} (2002) 78. 
\bibitem{armesto04} N.~Armesto, C.~A.~Salgado and U.~A.~Wiedemann, Phys.\ Rev.\ Lett.\  {\bf 94} (2005) 022002.  
\bibitem{phobos_wp}B. B. Back {\it et al.} [PHOBOS], \NPA 757 (2005) 28. 
\bibitem{brahms_wp}I. Arsene {\it et al.} [BRAHMS], \NPA 757 (2005) 1. 
\bibitem{tuchin04}D.Kharzeev, Y.Kovchegov, K.Tuchin, Phys. Lett. B{\bf 599} (2004) 23
\bibitem{hf}A.~S.~Ayan {\it et al.},  J.\ Phys.\ G {\bf 30} (2004) N33. 
\bibitem{totem}V.~Berardi {\it et al.}  [TOTEM], ``TOTEM: Technical design report'', CERN-LHCC-2004-002.
\bibitem{castor}P.~Katsas {\it et al.}, submitted to \NIMA.
\bibitem{zdc}O.~A.~Grachov {\it et al.}, Proceeds. CALOR'06, nucl-ex/0608052. 
\bibitem{mueller_navelet}A.~H.~Mueller and H.~Navelet,  Nucl.\ Phys.\ B {\bf 282} (1987) 727.  
\bibitem{dde_qm05}D.~d'Enterria, {\it Proceeds. Quark Matter'05}, nucl-ex/0601001. 
\bibitem{starlight}S.~R.~Klein, J.~Nystrand, \PRC 60 (1999) 014903; A.~Baltz, S.~Klein, J.~Nystrand, \PRL 89 (2002) 012301.
\end{thebibliography}
\end{document}